\documentclass[useAMS,usenatbib,usegraphicx]{mn2e}
\usepackage{color}
\usepackage{wasysym}
\usepackage{url}
\usepackage{multirow}
\usepackage{times}
\usepackage[usenames,dvipsnames]{xcolor}

\newcommand{\nuc}[2]{\ensuremath{\mathrm{^{#1}#2}}}
\newcommand{\ions}[2]{#1\,{\sc #2}}

\newcommand{\msun}{\ensuremath{\mathrm{M}_\odot}}
\newcommand{\gccm}{\ensuremath{\mathrm{g} \, \mathrm{cm}^{-3}}}
\newcommand{\mch}{\ensuremath{M_\mathrm{Ch}}}
\def\lesssim{\mathrel{\hbox{\rlap{\hbox{\lower4pt\hbox{$\sim$}}}\hbox{$<$}}}}

\def\gtrsim{\mathrel{\hbox{\rlap{\hbox{\lower4pt\hbox{$\sim$}}}\hbox{$>$}}}}

\def\aj{AJ}%
\def\araa{ARA\&A}%
\def\apj{ApJ}%
\def\apjl{ApJL}%
\def\apjs{ApJS}%
\def\aap{A\&A}%
\def\mnras{MNRAS}%
\def\pasp{PASP}%
\def\nat{Nature}%
\title[Deflagrations in hybrid CONe white dwarfs]
{%
  Deflagrations in hybrid CONe white dwarfs: a route to explain the faint
  Type Iax supernova 2008ha
}

\author[M.~Kromer~et~al.]
{M.~Kromer,$^{1}$\thanks{E-mail: markus.kromer@astro.su.se}
  S.~T.~Ohlmann,$^{2,3}$
  R.~Pakmor,$^{3}$
  A.~J.~Ruiter,$^{4,5}$
  W.~Hillebrandt,$^{6}$
  \newauthor
  K.~S.~Marquardt,$^{2,3}$
  F.~K.~R\"{o}pke$,^{3,7}$
  I.~R.~Seitenzahl,$^{4,5}$
  S.~A.~Sim$^{5,8}$
  and S.~Taubenberger$^{6,9}$\\
  $^{1}$The Oskar Klein Centre \& Department of Astronomy,
       Stockholm University, AlbaNova, SE-106 91 Stockholm, Sweden\\
  $^{2}$Institut f\"ur Theoretische Physik und Astrophysik, Universit\"at
        W\"urzburg, Emil-Fischer-Stra{\ss}e 31, D-97074 W\"urzburg, Germany\\
  $^{3}$Heidelberger Institut f\"{u}r Theoretische Studien, 
        Schloss-Wolfsbrunnenweg 35, D-69118 Heidelberg, Germany\\
  $^{4}$Research School of Astronomy \& Astrophysics, 
        Mount Stromlo Observatory, Cotter Road, Weston ACT 2611,
        Australia\\
  $^{5}$ARC Centre of Excellence for All-Sky Astrophysics (CAASTRO)\\
  $^{6}$Max-Planck-Institut f{\"u}r Astrophysik, 
       Karl-Schwarzschild-Str. 1, D-85748 Garching bei M{\"u}nchen, Germany\\
  $^{7}$Zentrum f{\"u}r Astronomie der Universit{\"a}t Heidelberg, 
       Institut f{\"u}r Theoretische Astrophysik, Philosophenweg 12, 
       D-69120 Heidelberg, Germany\\
  $^{8}$Astrophysics Research Centre, School of Mathematics and Physics, 
       Queen's University Belfast, Belfast BT7 1NN, UK\\
  $^{9}$European Southern Observatory, 
       Karl-Schwarzschild-Str. 2, D-85748 Garching bei M{\"u}nchen, Germany\\  
}

\begin{document}

\date{Accepted, 17 April 2015. Received, 14 March 2015; in original form, 14
March 2015}

\pagerange{\pageref{firstpage}--\pageref{lastpage}} \pubyear{2012}

\maketitle

\label{firstpage}

\begin{abstract}
  Stellar evolution models predict the existence of hybrid white
  dwarfs (WDs) with a carbon--oxygen core surrounded by an
  oxygen--neon mantle. Being born with masses $\sim 1.1\,\msun$,
  hybrid WDs in a binary system may easily approach the Chandrasekhar
  mass (\mch) by accretion and give rise to a thermonuclear
  explosion. Here, we investigate an off-centre deflagration in a
  near-\mch\ hybrid WD under the assumption that nuclear burning only
  occurs in carbon-rich material. Performing hydrodynamics simulations
  of the explosion and detailed nucleosynthesis post-processing
  calculations, we find that only 0.014\,\msun\ of material is ejected
  while the remainder of the mass stays bound. The ejecta consist
  predominantly of iron-group elements, O, C, Si and S. We also
  calculate synthetic observables for our model and find reasonable
  agreement with the faint Type Iax SN~2008ha. This shows for the
  first time that deflagrations in near-\mch\ WDs can in principle
  explain the observed diversity of Type Iax supernovae. Leaving
  behind a near-\mch\ bound remnant opens the possibility for
  recurrent explosions or a subsequent accretion-induced collapse in
  faint Type Iax SNe, if further accretion episodes occur. From binary
  population synthesis calculations, we find the rate of hybrid WDs
  approaching \mch\ to be on the order of 1 percent of the Galactic
  SN~Ia rate.
\end{abstract}


\section{Introduction}

Type Iax supernovae (SNe~Iax) form a distinct class of astronomical
transients \citep{foley2013b} with SN~2002cx \citep{li2003a,
  branch2004a, jha2006a} as the proto-typical example. It has been
argued that SNe~Iax occur at a rate of 5 to 30 percent of the overall
rate of Type Ia supernovae (SNe~Ia) \citep{li2011a, foley2013b,
  white2015a}.  Recently, \citet{mccully2014a} reported the first
detection of a progenitor candidate for a SN~Iax. In pre-explosion
\textit{Hubble Space Telescope (HST)} images they discovered a point
source at the location of SN~2012Z. Future \textit{HST} observations
will show whether this point source is actually a massive star that
exploded giving rise to the SN, or a He donor to an accreting
carbon--oxygen (CO) white dwarf (WD) as argued by
\citet{mccully2014a}.

Although spectroscopically similar to SNe~Ia, SNe~Iax are
characterized by distinctive features: (i) expansion velocities at
maximum light (2000 to 8000\,km\,s$^{-1}$) that are significantly
lower than those of SNe~Ia (10,000 to 15,000\,km\,s$^{-1}$, e.g.\
\citealt{foley2013b, stritzinger2015a}), (ii) low peak absolute
magnitudes with respect to the width--luminosity relation of SN~Ia
light curves \citep[e.g.][]{mcclelland2010a,stritzinger2014a}, (iii)
peculiar late-time spectra that differ strongly from other SNe
\citep[e.g.][]{jha2006a, sahu2008a} and (iv) a statistical association
with star-formation regions (so far no SN~Iax has been observed in
elliptical galaxies), indicating short delay times of 30 to 50\,Myr
\citep{lyman2013a}.

SNe~Iax show significant diversity in peak absolute magnitude and
expansion velocities. For example, SN~2005hk, one of the brighter
class members, peaked at $M_\mathrm{max}^V\sim-18.1$\,mag
\citep{chornock2006a, phillips2007a, sahu2008a}. In contrast,
SN~2008ha, to date the faintest member of the class, reached only
$M_\mathrm{max}^V\sim-14.2$\,mag \citep{foley2009a, valenti2009a,
  foley2010a}.  This indicates a wide range of ejecta properties.

Several explosion mechanisms have been proposed for SNe~Iax, including
core collapse of massive stars \citep{foley2009a, valenti2009a,
  moriya2010a}, detonations in He shells on top of CO WDs
\citep{foley2009a} and deflagrations or pulsational delayed
detonations in near Chandrasekhar-mass (\mch) CO WDs
\citep{branch2004a, phillips2007a, stritzinger2015a}. The mechanism
that seems to agree best with the observed properties of SNe~Iax,
including the possible detection of a stellar remnant for the faint
Type Iax SN~2008ha \citep{foley2014b}, is a deflagration in a
near-\mch\ CO WD that fails to completely unbind the WD, leaving
behind a bound remnant \citep{foley2009a}.

\citet{jordan2012b} were the first to present hydrodynamical
simulations of this scenario yielding \nuc{56}{Ni} and total ejecta
masses in agreement with the observationally derived quantities for
bright SNe~Iax.
Combining detailed explosion models and radiative transfer
simulations, \citet{kromer2013a} have shown that pure deflagrations in
near-\mch\ CO WDs that leave behind a compact remnant can indeed
reproduce the observational display of the brighter SNe~Iax like SNe
2002cx or 2005hk. \citet{fink2014a} have confirmed these results with
a set of models exploring deflagrations in near-\mch\ CO WDs for
different ignition setups. However, they fail to explain the faint
class members like SN~2008ha (their faintest model yields
$M_\mathrm{max}^V=-16.8$\,mag while SN~2008ha peaks at $-14.2$\,mag;
\citealt{foley2009a}). Although \citet{jordan2012b} and
\citet{fink2014a} did not fully explore the parameter space of
possible ignition configurations, it seems implausible to reach low
enough \nuc{56}{Ni} masses to explain the low luminosities of the
faintest SNe~Iax from deflagrations in near-\mch\ CO WDs.

To explain these events with deflagrations in near-\mch\ WDs, one
somehow has to quench burning at an early stage to reduce the
\nuc{56}{Ni} production. This is potentially possible by an abrupt
change in the composition profile of the exploding WD as seen in
hybrid WDs where a CO core is surrounded by an oxygen--neon (ONe)
mantle. Such objects have recently been predicted by stellar evolution
models \citep{denissenkov2013c, chen2014a}. \citet{denissenkov2015a}
have shown that a thermonuclear runaway in accreting hybrid CONe WDs
is possible when they approach \mch.  Since hybrid CONe WDs originate
from relatively massive zero-age main sequence (ZAMS) stars
\citep{chen2014a}, such a scenario could also explain the short delay
times of SNe~Iax \citep{meng2014a}.

Here, we study a deflagration in such a near-\mch\ hybrid CONe WD.
The paper is organised as follows: in Section~\ref{sec:progenitor} we
discuss the chosen progenitor model before we present our explosion
simulation in Section~\ref{sec:explosion}. Synthetic observables for
our model and comparison to observational data are presented in
Section~\ref{sec:obscomp}. In Section~\ref{sec:discussion} we discuss
our results and constraints on rates and delay times from binary
population synthesis calculations before presenting our conclusion in
Section~\ref{sec:conclusion}.

\section{Progenitor model}
\label{sec:progenitor}

In the single-degenerate progenitor scenario for SNe~Ia, a CO WD is
supposed to accrete H-rich material from a binary companion until it
approaches \mch\ and a thermonuclear runaway occurs \citep[e.g.][for a
review]{hillebrandt2000a}. CO WDs are formed when the cores of
asymptotic giant branch (AGB) stars are exposed after they have
expelled their envelopes.  In classical single-star evolution models
the maximum mass for the CO core of an AGB star is $\sim 1.05$\,\msun\
\citep{chen2014a}. For more massive cores, C burning is ignited
(super-AGB phase) and the entire core is converted to O and Ne,
leading to the formation of an ONe WD once the star has lost its
envelope. ONe WDs have been discarded as progenitor candidates for the
single-degenerate progenitor scenario of SNe~Ia. It has been argued
that they should undergo a gravitational collapse and form a neutron
star rather than a thermonuclear explosion when approaching \mch\
\citep{nomoto1984b,nomoto1987a,nomoto1991a}.

Recently, \citet{denissenkov2013c} and \citet{chen2014a} have shown
that convective boundary mixing during the super-AGB phase can prevent
off-centre ignited C burning from reaching the centre of the star,
thus leaving a hybrid CONe WD with a C-rich core at the end of the
super-AGB phase. In contrast to an ONe WD, such a hybrid WD with a
C-rich core is likely to be able to trigger a thermonuclear explosion
when approaching \mch.  ONe-rich WDs containing a small fraction of
unburnt C have been predicted by stellar evolution models previously
\citep{siess2006a}, and have been mentioned in the literature as
possible progenitors of thermonuclear explosions
\citep{garcia-berro1997a}.

\citet{denissenkov2015a} have simulated the mass accretion onto hybrid
CONe WDs to investigate their potential as SN~Ia progenitors in the
single-degenerate scenario. They find that a thermonuclear runaway in
accreting hybrid CONe WDs is possible depending on the mass of the
parent star and accretion history. The exact thermal and chemical
structure of the hybrid WD at the time of explosion, however, is
uncertain and can only be determined from 3D reactive-convective hydro
simulations. In particular, a series of convective Urca shell flashes
can either limit the mass of the convective C core or completely
suppress convective mixing \citep{denissenkov2015a}.

Here, we investigate how a deflagration will incinerate a hybrid CONe
WD that undergoes a thermonuclear runaway when approaching \mch\ and
calculate synthetic observables to compare the optical display of such
an explosion to observed SNe~Ia(x). For this purpose we neglect the
uncertainties regarding the exact structure of the progenitor and
follow a simplified approach to model the hybrid CONe WD. Motivated by
the model of \citet{denissenkov2013c}, we choose the initial WD
(central density $\rho_\mathrm{c}=2.9\times10^9\,\gccm$ and isothermal
temperature structure with $T=5\times10^5$\,K) to have a C-rich core
of $0.2 M_\odot$ ($X(\mathrm{C})=0.5$, $X(\mathrm{O})=0.5$),
surrounded by an ONe layer ($X(\mathrm{C})=0.03$, $X(\mathrm{O})=0.5$,
$X(\mathrm{Ne})=0.47$) up to 1.1\,\msun. On top of this, we place a
layer of accreted material with $X(\mathrm{C})=0.5$,
$X(\mathrm{O})=0.5$ which extends up to 1.4\,\msun.

\section{Explosion simulation}
\label{sec:explosion}
To simulate the explosion phase we use our 3D hydro code
\textsc{leafs} \citep[based on the implementation
by][]{reinecke2002b}. \textsc{leafs} solves the reactive Euler
equations and models the propagation of deflagrations using a levelset
scheme \citep{reinecke1999a, osher1988a, smiljanovski1997a} and
sub-grid scale turbulence model \citep{schmidt2006b,schmidt2006c} to
account for flame acceleration by buoyancy- and shear-instability
induced turbulent fluid flows. Self-gravity is dealt with by a
monopole gravity solver and an equation of state appropriate for WD
matter is used. To follow the evolution of the explosion ejecta to the
phase of homologous expansion, we use the expanding grid technique
described by \citet{roepke2005c} and \citet{roepke2006a}. For more
information on the hydrodynamics code and details concerning the
burning treatment see \citet{fink2014a}. 

\begin{figure*}
  \centering
  \includegraphics[width=\linewidth]{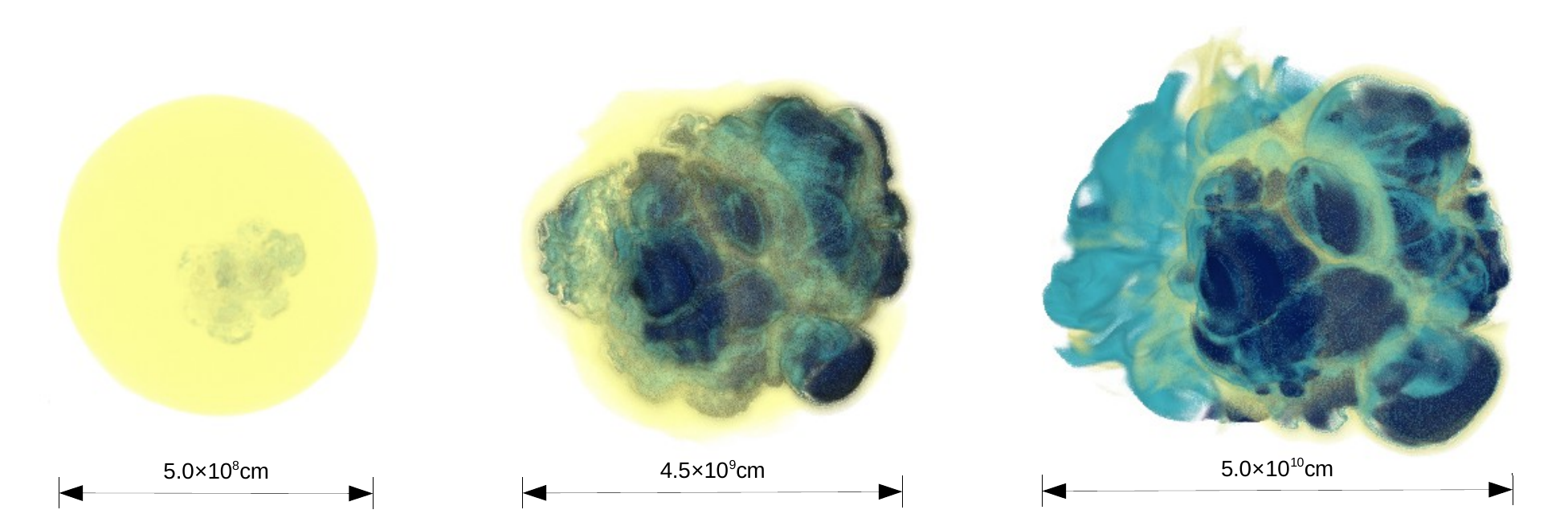}
  \caption{Hydrodynamical evolution of our model at 0.85, 2.5 and
    10\,s after explosion (from left to right). We show a volume
    rendering of the mean atomic number where yellow corresponds to
    unburnt material, and cyan and dark blue to ejecta rich in
    intermediate-mass or iron-group elements, respectively.}
  \label{fig:explosion}
\end{figure*}

In our progenitor WD as described in Section~\ref{sec:progenitor} we
ignite a deflagration near the centre using the same asymmetric
5-kernel ignition configuration as in the N5def model of
\citet{kromer2013a} and \citet{fink2014a}. This model of the
deflagration of a near-\mch\ CO WD produces observables in good
agreement with the Type Iax SN~2005hk \citep{kromer2013a}. Here, we
follow the flame propagation in our hybrid CONe WD and the subsequent
hydrodynamical evolution of the ejecta up to the phase of homologous
expansion (see Figure~\ref{fig:explosion}).

After the burning starts from the 5 ignition kernels, the flames soon
merge and a one-sided deflagration plume develops within the C-rich
core. In the ONe layer above $0.2\,\msun$, it is assumed that the C
deflagration flame cannot propagate because of the small C fraction
($X(\mathrm{C})=0.03$).  Thus, burning is switched off in the region
initially consisting of mainly ONe matter and further out. Burning
ceases at $\sim1.5$\,s past explosion. The hot ashes rise to the
surface and wrap around the unburnt parts of the progenitor WD. 
Due to the small nuclear energy released in the burning
($E_\mathrm{nuc}=1.0\times10^{50}$\,erg), only $0.014\,\msun$ of
material is ejected while a massive bound remnant of $1.39\,\msun$ is
left behind (the mass of the bound remnant and ejecta are determined
as described by \citealt{kromer2013a}). We stop the simulation at
100\,s after ignition. At that time the ejecta are in homologous
expansion with an asymptotic kinetic energy of
$E_\mathrm{kin}=1.8\times10^{48}$\,erg.

In a postprocessing calculation with a $384$-isotope nuclear network
\citep{travaglio2004a} we determine the detailed chemical composition
of the ejecta. To this end, about $10^6$ Lagrangian tracer particles
are passively advected in the hydrodynamic explosion simulation and
thermodynamic trajectories are recorded for each tracer particle. The
tracers are placed in the WD with variable masses according to the
prescription by \citet{seitenzahl2010a}. The initial composition of
the tracer particles is drawn from the abundance profile of our
progenitor WD. 

Within the ejecta, we find $6.41\times10^{-3}$\,\msun\ of iron-group
elements (IGEs), of which $3.40\times10^{-3}$\,\msun\ are
\nuc{56}{Ni}. The most abundant other species are O
($3.49\times10^{-3}$\,\msun), C ($2.73\times10^{-3}$\,\msun), Si
($8.02\times10^{-4}$\,\msun) and S ($3.56\times10^{-4}$\,\msun). The
detailed composition of the ejecta is given in
Tables~\ref{tab:yields} and \ref{tab:ryields}.

\begin{table}
  \caption{Asymptotic chemical yields in the ejecta of our model.}
  \label{tab:yields}
  \centering
  \begin{tabular}{lcc}
    \hline
     & \msun \\
    \hline
Total & $ 1.40\times10^{-2}$\\
 He &  $ 3.58\times10^{-7}$\\
 Li &  $ 3.90\times10^{-13}$\\
 Be &  $ 4.58\times10^{-13}$\\
  B &  $ 1.83\times10^{-11}$\\
  C &  $ 2.73\times10^{-3}$\\
  N &  $ 1.79\times10^{-7}$\\
  O &  $ 3.49\times10^{-3}$\\
  F &  $ 2.58\times10^{-9}$\\
 Ne &  $ 3.88\times10^{-4}$\\
 Na &  $ 1.75\times10^{-6}$\\
 Mg &  $ 7.60\times10^{-5}$\\
 Al &  $ 3.78\times10^{-6}$\\
 Si &  $ 8.02\times10^{-4}$\\
  P &  $ 1.95\times10^{-6}$\\
  S &  $ 3.56\times10^{-4}$\\
    \hline
  \end{tabular}
  \begin{tabular}{lcc}
    \hline
     & \msun \\
    \hline
 Cl &  $ 1.00\times10^{-6}$\\
 Ar &  $ 6.29\times10^{-5}$\\
  K &  $ 6.98\times10^{-7}$\\
 Ca &  $ 4.74\times10^{-5}$\\
 Sc &  $ 3.24\times10^{-9}$\\
 Ti &  $ 1.68\times10^{-6}$\\
  V &  $ 9.35\times10^{-7}$\\
 Cr &  $ 8.71\times10^{-5}$\\
 Mn &  $ 1.57\times10^{-4}$\\
 Fe &  $ 5.28\times10^{-3}$\\
 Co &  $ 9.05\times10^{-6}$\\
 Ni &  $ 8.76\times10^{-4}$\\
 Cu &  $ 5.74\times10^{-8}$\\
 Zn &  $ 6.15\times10^{-8}$\\
 Ga &  $ 1.20\times10^{-9}$\\
 Ge &  $ 1.63\times10^{-9}$\\
    \hline
  \end{tabular}
\end{table}

\begin{table}
  \caption{Yields of radioactive isotopes in the model ejecta at $t=100$\,s.}
  \label{tab:ryields}
  \centering
  \begin{tabular}{lc}
    \hline
     & \msun \\
    \hline
\nuc{56}{Ni}	& $3.40\times10^{-3}$\\
\nuc{57}{Ni}	& $1.29\times10^{-4}$\\
\nuc{55}{Co}	& $1.18\times10^{-4}$\\
\nuc{52}{Fe}	& $4.36\times10^{-5}$\\
\nuc{55}{Fe}	& $3.62\times10^{-5}$\\
\nuc{57}{Co}	& $1.82\times10^{-5}$\\
\nuc{53}{Fe}	& $7.84\times10^{-6}$\\
\nuc{59}{Ni}	& $7.56\times10^{-6}$\\
\nuc{53}{Mn}	& $3.96\times10^{-6}$\\
\nuc{56}{Co}	& $2.21\times10^{-6}$\\
\nuc{48}{Cr}	& $1.17\times10^{-6}$\\
\nuc{51}{Mn}	& $6.28\times10^{-7}$\\
\nuc{37}{Ar}	& $1.58\times10^{-7}$\\
\nuc{60}{Fe}	& $1.48\times10^{-7}$\\
\nuc{62}{Zn}	& $1.33\times10^{-7}$\\
\nuc{58}{Co}	& $1.21\times10^{-7}$\\
\nuc{54}{Mn}	& $1.12\times10^{-7}$\\
\nuc{49}{Cr}	& $1.11\times10^{-7}$\\
\nuc{14}{C}	& $1.07\times10^{-7}$\\
\nuc{51}{Cr}	& $9.11\times10^{-8}$\\
    \hline
  \end{tabular}
  \begin{tabular}{lc}
    \hline
     & \msun \\
    \hline
\nuc{59}{Fe}	& $8.15\times10^{-8}$\\
\nuc{52}{Mn}	& $4.95\times10^{-8}$\\
\nuc{41}{Ca}	& $4.16\times10^{-8}$\\
\nuc{44}{Ti}	& $3.95\times10^{-8}$\\
\nuc{60}{Co}	& $3.02\times10^{-8}$\\
\nuc{63}{Ni}	& $1.79\times10^{-8}$\\
\nuc{26}{Al}	& $1.34\times10^{-8}$\\
\nuc{35}{S}	& $2.73\times10^{-9}$\\
\nuc{65}{Zn}	& $2.42\times10^{-9}$\\
\nuc{49}{V}	& $2.33\times10^{-9}$\\
\nuc{39}{Ar}	& $2.22\times10^{-9}$\\
\nuc{32}{P}	& $2.12\times10^{-9}$\\
\nuc{36}{Cl}	& $2.00\times10^{-9}$\\
\nuc{33}{P}	& $1.60\times10^{-9}$\\
\nuc{48}{V}	& $1.00\times10^{-9}$\\
\nuc{40}{K}	& $6.25\times10^{-10}$\\
\nuc{22}{Na}	& $3.32\times10^{-10}$\\
\nuc{32}{Si}	& $2.81\times10^{-10}$\\
\nuc{68}{Ge}	& $7.98\times10^{-11}$\\
\nuc{65}{Ga}	& $1.43\times10^{-11}$\\
    \hline
  \end{tabular}

\end{table}

\begin{figure*}
  \centering
  \includegraphics[width=0.9\linewidth]{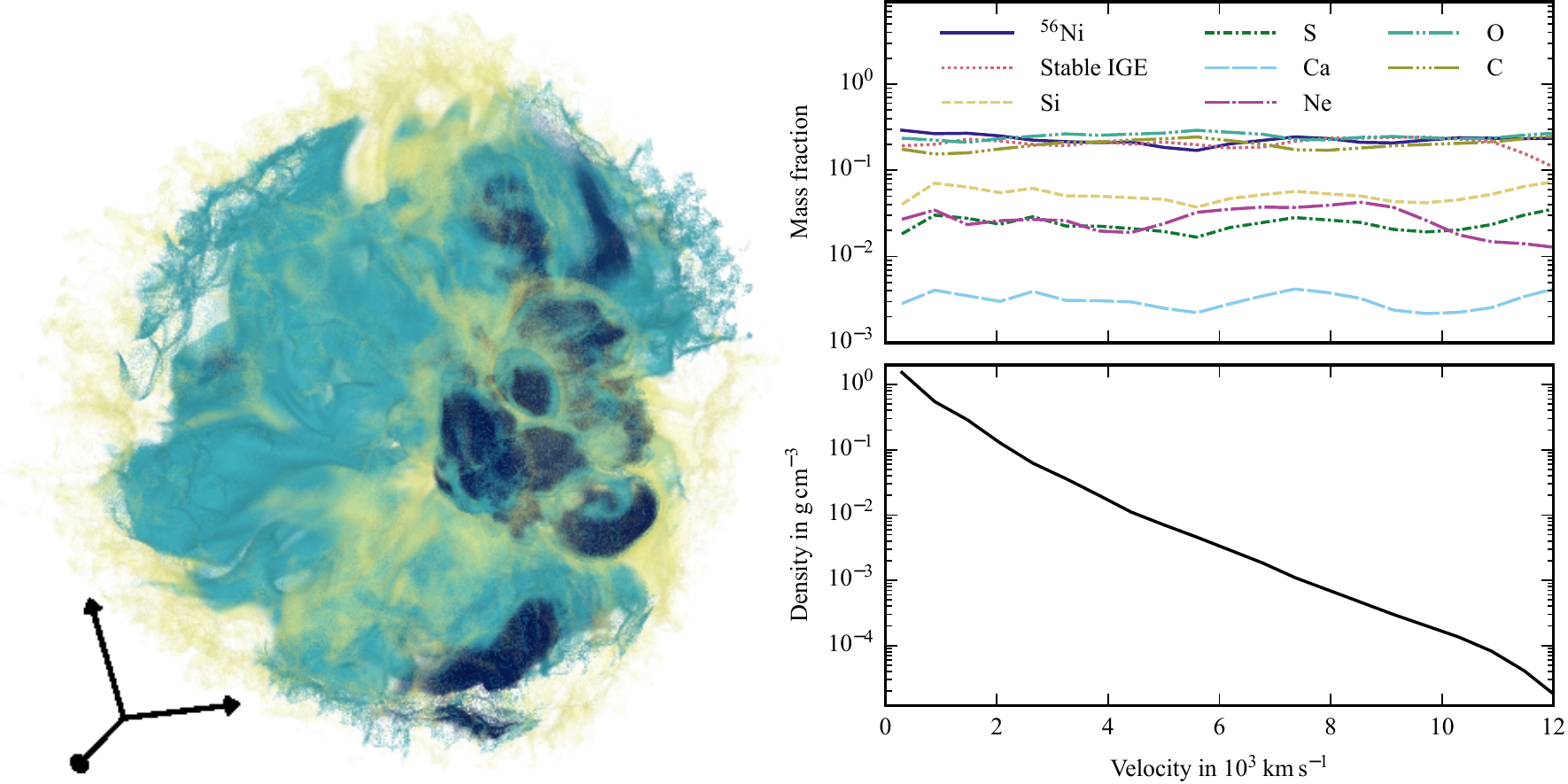}
  \caption{Final ejecta structure at 100\,s. The left-hand panel
    shows a 3D volume rendering of the mean atomic number as in
    Figure~\ref{fig:explosion} (the triad corresponds to a length of
    $5\times10^{10}$\,cm). The right-hand panel shows angle-averaged
    density and mass fractions of selected species as a function of
    expansion velocity.}
  \label{fig:composition}
\end{figure*}

Using an SPH-like algorithm \citep{kromer2010a} we map the final
composition of the unbound tracer particles to a $200^3$ Cartesian
grid to determine the abundance structure of the ejecta (see
Figure~\ref{fig:composition}).  Our model shows significant deviations
from spherical symmetry which is a consequence of the turbulent
burning and the one-sided propagation of the thermonuclear flame
within the progenitor WD.

For the bound remnant, which is not resolved in the hydrodynamics
calculations, our nucleosynthesis post-processing is not reliable,
since the tracers do not record accurate thermodynamic trajectories.
Thus, we can only give approximate yields from the simplified
description of nuclear reactions implemented in our hydro scheme. From
this we find 1.33\,\msun\ unburnt material, $3.3\times10^{-3}$\,\msun\
intermediate-mass elements and $5.3\times10^{-2}$\,\msun\ IGEs, of
which $2.0\times10^{-2}$\,\msun\ are \nuc{56}{Ni}.

\section{Comparison with observations}
\label{sec:obscomp}

Using our 3D Monte Carlo radiative transfer code \textsc{artis}
\citep{kromer2009a,sim2007b} we calculate synthetic observables for
the ejecta. To this end we re-map the abundance and density structure
as obtained from the explosion simulation to a $50^3$ Cartesian
grid. On this grid we follow the propagation of $1.2 \times 10^8$
Monte Carlo quanta for 136 logarithmic time steps from $0.3$ to
$35$\,d past explosion.  Escaping Monte Carlo packets are binned in
time and on a logarithmic wavelength grid spanning 10,000 bins from
600 to 30,000\,\AA\ to obtain a spectral time sequence.  For our
simulation we use the atomic data set as described by
\citet{gall2012a}.  Local thermodynamic equilibrium is assumed for $t
< 0.42$\,d and a grey approximation is used in optically thick cells
\citep[cf.][]{kromer2009a} to speed up the calculations.

\begin{figure*}
  \centering
  \includegraphics[width=0.9\linewidth]{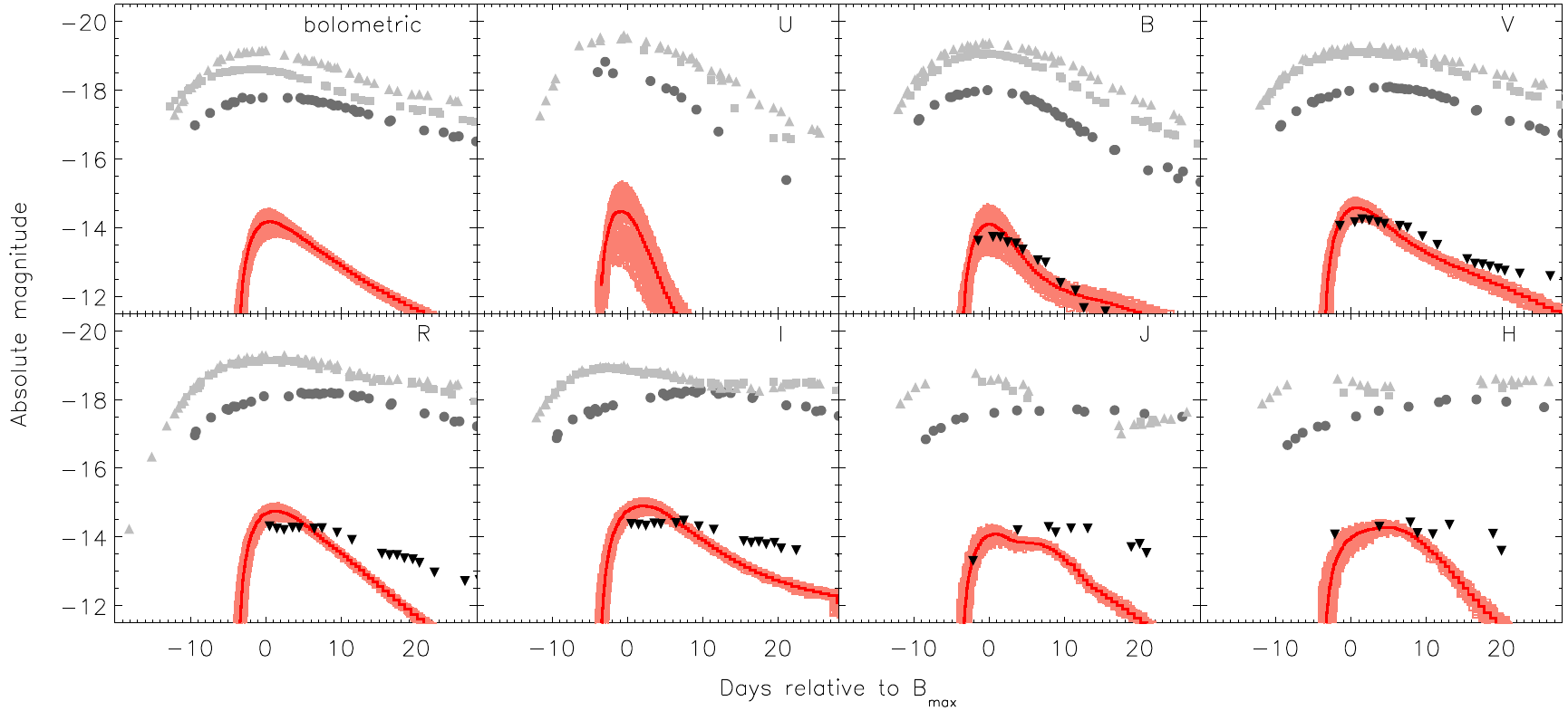}
  \caption{Synthetic light curves of our model for different filter
    bands as indicated in the panels. We show light curves from 100
    different viewing angles (light red) sampling the full range of
    solid angle and the angle average (dark red). For comparison we
    show observational data for the normal SNe~Ia 2004eo, 2005cf (grey
    squares and triangles, respectively), SN~2005hk (a proto-typical
    SN~Iax, dark grey circles) and SN~2008ha (black upside-down
    triangles). The latter marks the faint end of the current sample
    of SNe~Iax.}
  \label{fig:lightcurves}
\end{figure*}

Synthetic broad-band light curves of our model are shown in
Figure~\ref{fig:lightcurves}. To examine the effects of asymmetries in
the explosion ejecta (cf. Figure~\ref{fig:composition}), we extract
light curves along 100 different viewing angles sampling the full
solid angle. The peak magnitudes of our model span a range from
$-13.2$ to $-14.6$ and $-14.2$ to $-14.8$ in $B$ and $V$ band,
respectively. These are significantly fainter than normal SNe~Ia or
proto-typical SNe~Iax like SN~2002cx or SN~2005hk. Our peak magnitudes
are, however, in good agreement with those of the faint Type Iax
SN~2008ha ($M_\mathrm{max}(B)=-13.74$\,mag,
$M_\mathrm{max}(V)=-14.21$\,mag; \citealt{foley2009a}). We also find
reasonable agreement for the maximum-light colours along selected
lines-of-sight.

However, our model light curves evolve significantly faster than
SN~2008ha. This becomes obvious comparing the rise times of the model
($t_\mathrm{max}(B)=2.9 \dots 4.6$\,d depending on the viewing angle)
and SN~2008ha ($\sim$10\,d; \citealt{foley2009a}), and also the
decline of the redder bands at epochs later than $\sim$10\,d past
$B$-band maximum. The fast evolution of our model is a consequence of
the extremely low ejecta mass of $0.014\,\msun$, which is
significantly smaller than the ejecta mass derived for
SN~2008ha. Using analytical light curve models \citet{foley2009a}
inferred an ejecta mass of $M_\mathrm{ej}=0.15\,\msun$ based on a
first set of post-maximum observations. Including a pre-maximum
spectrum to the analysis \citet{foley2010a} revised this estimate to
$M_\mathrm{ej}=0.30$\,\msun\ (\citealt{valenti2009a} obtain an ejecta
mass in the range 0.1 to 0.5\,\msun).

As a consequence of the relatively strong asymmetry of the model
ejecta, our light curves show stronger sensitivity to different
viewing angles than those of the N5def model \citep{kromer2013a}. At
peak the $U$-, $B$- and $V$-band magnitudes show a scatter of 2.3, 1.4
and 0.6 mag, respectively (Figure~\ref{fig:lightcurves}). Given the
small sample size of faint SNe~Iax and the large intrinsic diversity
in the few systems known, this viewing angle sensitivity cannot be
used to rule out the model.

\begin{figure}
  \centering
  \includegraphics[width=0.9\linewidth]{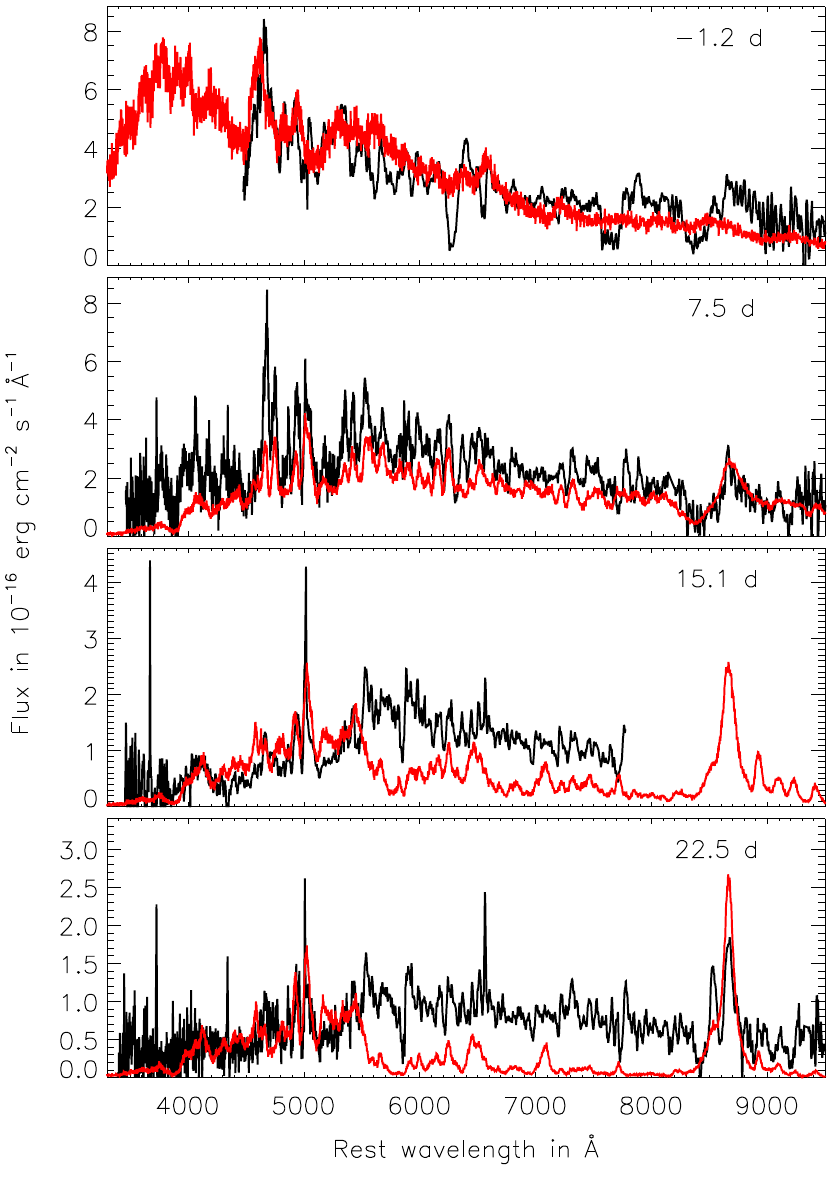}
  \caption{Angle-averaged synthetic spectra (in red) of our model at
    different epochs (times with respect to $B$-band maximum are
    indicated by the labels in the panels). For comparison, observed
    spectra of SN~2008ha \citep{foley2009a,foley2010a,valenti2009a}
    are shown (the flux calibration has been checked against the
    photometry and adjusted if necessary). The observed spectra have
    been de-reddened and de-redshifted.}
  \label{fig:spectra}
\end{figure}

Selected snapshots of the synthetic spectral time series resulting
from our simulations are shown in Figure~\ref{fig:spectra} and compared
to observed spectra of SN~2008ha at corresponding epochs. At $-1.2$\,d
(with respect to $B$-band maximum) our model agrees fairly well with
the spectral energy distribution of SN~2008ha. However, the model
fails to reproduce some of the narrow line features present in
SN~2008ha, e.g.\ the features at 6300\,\AA\,, 7600\,\AA\,, and 8400\,\AA\
that have previously been identified with \ions{Si}{ii}, \ions{O}{i}
and \ions{Ca}{ii}, respectively \citep{foley2010a}. This could
indicate a lack of intermediate-mass elements in the outer layers of
the model ejecta. 

Another explanation for the lack of narrow features in our model could
be too large ejecta velocities. The model spectra are dominated by
fluorescent emission in a plethora of atomic lines of IGEs, which are
present throughout the ejecta in our model (see
Figure~\ref{fig:composition}).  As long as the emission originates
from the outer layers (where the expansion velocity is large) the
emission lines are blended giving rise to a smooth pseudo continuum.
With time the ejecta expand further, the outer layers become
increasingly optically thin and deeper layers at lower velocities
start to dominate the emission. Then individual lines de-blend and
narrow features appear. A larger ejecta mass at similar energy release
could reduce the ejecta velocities and thus lead to narrower line
features.

\begin{figure}
  \centering
  \includegraphics[width=0.9\linewidth]{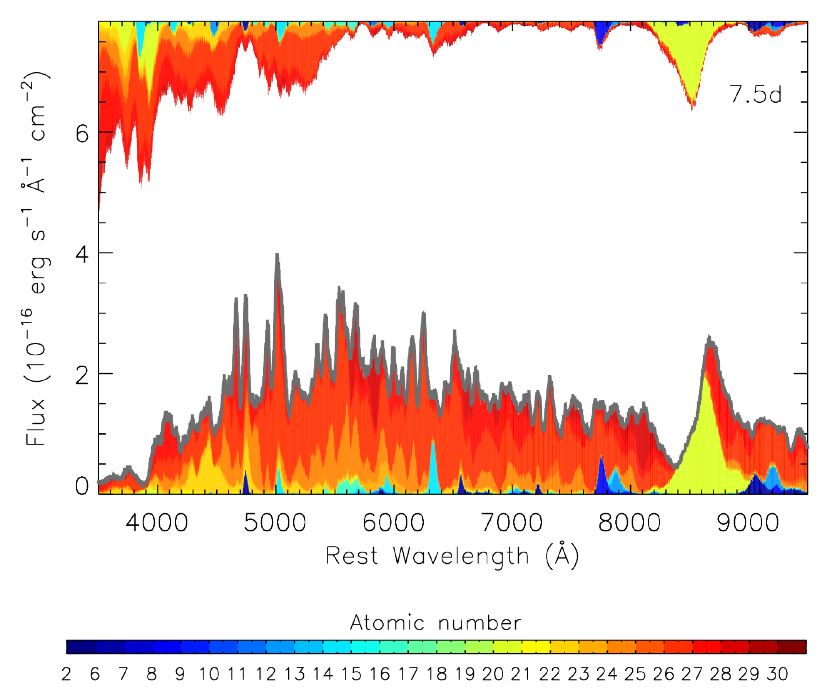}
  \caption{Angle-averaged synthetic spectrum of our model (grey) at
    7.5\,d past $B$-band maximum. The colour coding shows the chemical
    species responsible for both bound--bound emission and absorption
    of quanta in our Monte Carlo simulation.  Below the synthetic
    spectrum, we colour code the fraction of escaping quanta in each
    wavelength bin that were last emitted by a particular chemical
    species (corresponding atomic numbers are illustrated in the
    colour bar). The coloured regions along the top of the plot
    indicate which elements were last responsible for removing quanta
    from a particular wavelength bin (either by absorption or
    scattering\,/\,fluorescence).}
  \label{fig:specelements}
\end{figure}

A week after $B$-band maximum (7.5\,d) the agreement between our model
and SN~2008ha is fairly good. The model shows a forest of narrow line
features similar to that observed in SN~2008ha though not every
feature is reproduced perfectly. However, this cannot be expected from
a model that is not tuned to fit the data. Even for normal SNe~Ia,
``first-principle'' models do not perfectly reproduce the observed
spectra \citep[e.g.][]{sim2013a}, particularly if plotted in
\textit{absolute} fluxes as in Figure~\ref{fig:spectra}. In previous
work the narrow line features were interpreted as P-Cygni absorption
profiles (e.g.\ \citealt{foley2009a}). In our model most of the
features can be attributed to fluorescent emission in lines of IGEs
(see Figure~\ref{fig:specelements}).

At later epochs (15.1\,d and 22.5\,d) the agreement deteriorates. In
particular, the rapid decrease of flux at wavelengths larger than
5500\,\AA\ is not observed in SN~2008ha. This reflects the rapid
decline in some of our broad-band light curves ($R$ band and redder)
discussed above and indicates that the model ejecta become optically
thin too early, again pointing at a lack of ejecta mass in our model
compared to SN~2008ha. At late epochs, however, some approximations on
the plasma state, made in our radiative transfer code, become less
applicable \citep{kromer2009a}. Indeed, we find that the agreement
between synthetic observables and data deteriorates also for other
model classes \citep[e.g.][]{roepke2012a}. Thus, some of the
differences at later epochs could also be due to systematics in the
radiative transfer.

\section{Discussion}
\label{sec:discussion}

\subsection{Ejecta mass}

The lack of narrow line features around maximum light and the overly
rapid light curve evolution of our model compared to SN~2008ha
indicate that the ejecta mass of the model is too low. In contrast,
the good agreement in peak luminosity suggests that the \nuc{56}{Ni}
yield in the model ejecta is close to that in SN~2008ha. Here, we have
investigated a single realisation of a deflagration in a hybrid CONe
WD for a particular progenitor structure and ignition setup. In
reality both of these will show some diversity. The progenitor
structure will depend on the accretion history and the final
reactive-convective evolution of the WD core before the thermonuclear
runaway, which will also determine the ignition setup. A thorough
investigation of the parameter space is necessary to assess whether
variations in these initial conditions can increase the ejecta mass
while keeping the \nuc{56}{Ni} mass in the ejecta similar, and thus
lead to a better agreement with SN~2008ha.  This requires a large
number of explosion simulations and will be addressed in future
work. Given the large uncertainty in the chemical and thermal
structure of hybrid WDs \citep{denissenkov2015a}, here we focus on a
simple model for an exploratory study.

\subsection{Bound remnant}

A key prediction of our model is a bound stellar remnant of
$\sim\mch$. Recently, deep post-explosion \textit{HST} images have
revealed a point source S1 at the site of the faint Type Iax SN~2008ha
\citep{foley2014b}. One possible interpretation of S1, proposed by
\citet{foley2014b}, is that it could be an inflated stellar remnant of
a failed deflagration. From our present simulations, which follow the
evolution of the explosion ejecta on an expanding grid, we cannot
resolve the structure and thermal state of the remnant. Thus we cannot
compare our bound remnant to the observed properties of the point
source. We note, however, that explosion energy deposited in the bound
remnant could lead to an expanded envelope. Whether the expansion is
sufficient to account for a radius of $1500\,R_{\odot}$ as derived for
S1 \citep{foley2014b} remains to be seen. This question will be
addressed in more detail with an adaptive grid code in future work.

\begin{figure}
  \centering
  \includegraphics[width=0.9\linewidth]{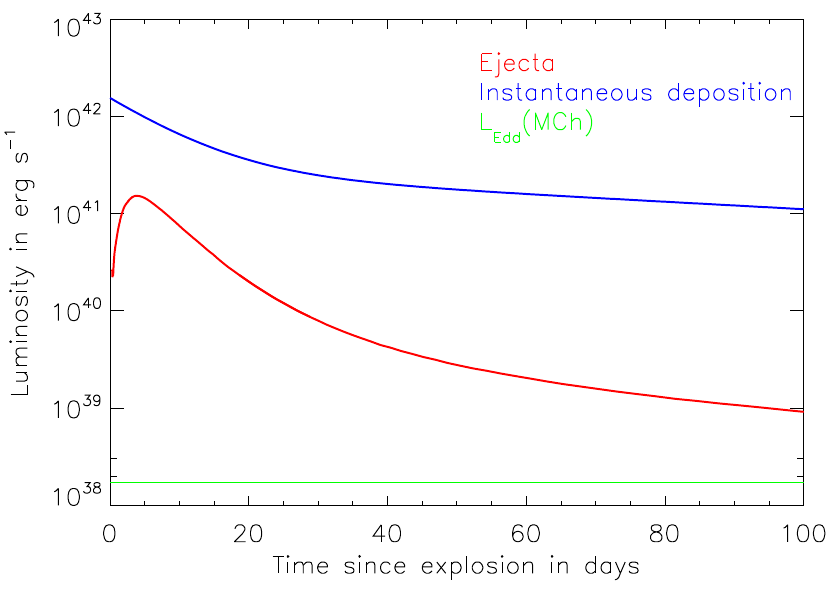}
  \caption{Bolometric light curve of the ejecta (red) and
    instantaneous energy deposition due to radioactive decay of
    \nuc{56}{Ni} in the bound remnant (blue). For comparison the green
    line shows the Eddington luminosity of a \mch\ object. To obtain
    the bolometric light curve of the ejecta for the full time range,
    we have performed an additional low-resolution radiative transfer
    simulation from 2 to 120\,d to extend our high-resolution
    simulation that stops at 35\,d.}
  \label{fig:late-lc}
\end{figure}

Moreover, our simulations show that $2.0\times10^{-2}$\,\msun\ of
\nuc{56}{Ni} synthesized during the explosion stay in the bound
remnant (see Section~\ref{sec:explosion}). This is $\sim 6$ times
larger than the \nuc{56}{Ni} yield in the ejecta. Consequently, the
instantaneous energy deposition by decaying \nuc{56}{Ni} in the bound
remnant is significantly larger than the bolometric luminosity of the
ejecta derived from our radiative transfer simulations (see
Figure~\ref{fig:late-lc}). Again, our ignorance of the structure of
the bound remnant makes it difficult to make a robust statement of how
the decay energy will affect the observational display. If the
\nuc{56}{Ni} is located close to the surface of the remnant, it might
drive further mass ejection and contribute significantly to the
observed luminosity. In contrast, if most of the \nuc{56}{Ni} is
confined to the central regions of the stellar remnant, it might take
a long time for the decay-deposited energy to diffuse to the surface
of the remnant so that it becomes negligible for the observational
display of the explosion.

Another interesting question concerns the long-term evolution of the
bound remnant. Because the kinetic energy of the explosion is very
low, the donor star is presumably not much affected by the impact of
the ejecta \citep{liu2013c}. Thus, it is possible that accretion
resumes once the remnant has relaxed into an equilibrium state. Since
only a tiny fraction of the mass of the initial WD is ejected, it is
likely that the bound remnant approaches the \mch\ limit
again. Whether or not this may lead to subsequent deflagrations and
thus a recurrent nature of some SNe~Iax, depends on the chemical and
thermal structure of the bound remnant. If the C fraction in the bound
remnant is low, an accretion-induced collapse and a neutron star might
be the more probable outcome. As our current simulations are unable to
resolve the bound remnant, we can not make a predictive statement
about its final fate at this stage.

\subsection{Burning in the ONe layer}

In this work, we have assumed that deflagration burning ceases when
the flame front reaches the ONe mantle of the hybrid WD. This
assumption significantly reduces the energy release compared to
deflagrations in CO WDs and is critical to obtain very low
\nuc{56}{Ni} masses as observed in faint SNe~Iax. However, in
principle, deflagrations are possible in ONe material as well
\citep{timmes1992a}. Since the energy release from burning this fuel
is lower, the flame slows down and its width increases more rapidly
with lower fuel density than for deflagrations in CO material
\citep{timmes1992a}. ONe deflagrations are therefore not expected to
propagate at densities lower than $\sim 10^9$\,\gccm. In our model,
the CO deflagration reaches the ONe layer when it has expanded to
densities of $\sim1.2\times10^9$\,\gccm, so some additional burning in
the ONe material is possible. Detailed microscopic flame simulations
are necessary, however, to assess our assumption that burning stalls
shortly after the deflagration reaches the ONe layer.

\subsection{Binary population synthesis - rates and constraints on
  delay times}

It is important to assess the likelihood of such events from a
theoretical standpoint: how frequent are the faint SN~Iax events?  We
evolved 300,000 binaries with Z=0.02 from the ZAMS up to a Hubble time
assuming a binary fraction of 70 percent using the population
synthesis code \textsc{StarTrack}
\citep{belczynski2002a,belczynski2008a}.  To obtain theoretical
birthrates we first calculated the number of ONe WDs that
approach \mch\ via stable Roche-lobe overflow from a stellar companion
\citep[see P-MDS model description in][]{ruiter2014a}. Typically,
these systems are considered to lead to accretion-induced collapse and
form neutron stars, but as discussed in Section~\ref{sec:progenitor},
if these WDs contain some non-negligible fraction of unburnt C, they
may instead lead to thermonuclear explosions.

The \textsc{StarTrack} code currently does not account for the
evolution of hybrid CONe WDs explicitly. However, it is reasonable to
assume that if such hybrid WDs exist then the lower mass limit will
occur near the boundary where, in our population models, a degenerate
CO core is formed, and where C burning occurs non-explosively leading
to the formation of a degenerate ONe core (the CO WD -- ONe WD
boundary; see \citet{belczynski2008a}). The upper limit for the hybrid
core mass will lie somewhere within the range of masses that are
canonically assumed to result in `pure' ONe WDs.

\citet{denissenkov2015a} found that single stars with ZAMS masses
between $\sim 6.4$ and $7.3$\,\msun\ produce CONe hybrid WDs. This
same mass range cannot be extrapolated to interacting binary stars
since a star that has lost or gained mass will follow a different
course of evolution (and end up with a different core mass) than that
of a single star with the same ZAMS mass.  To estimate how many of our
ONe WDs may contain some fraction of unburnt C in their cores, we
checked the corresponding WD birth masses that arise from ZAMS single
stars within this mass range in \textsc{StarTrack}. The corresponding
range is $1.193$ to $1.325$\,\msun.  Here, we assume these (ONe-rich)
WDs contain some fraction of unburnt C and thus are hybrid WDs. A
small number of ONe WDs are found below this mass boundary in our
model, and so we include these as potential hybrid cores as well. We
assume all of these WDs undergo an off-centre deflagration once they
approach \mch.

In terms of relative rates for different SN~Ia progenitors, we find
that over a Hubble time, hybrid CONe WDs that may produce faint
Iax-like events are 1 percent of the rate of the entire CO-CO WD
merger population. By comparison, they have about the same relative
rate that we predict for the classic single-degenerate scenario at
near solar metallicity, whereby a CO WD accretes toward \mch\ from a
hydrogen-burning star (cf.\ table 1 of Marquardt et al.,
submitted). To put it in a more absolute context: \citet{badenes2012a}
quote a SN~Ia rate of $1.1 \times 10^{-13}$\,yr$^{-1}$\,\msun$^{-1}$
for Milky Way like galaxies. We find from our population synthesis
model that the CO-CO WD merger rate (averaged over a Hubble time) is
$1.06 \times 10^{-13}$\,yr$^{-1}$\,\msun$^{-1}$, in other words: very
close to the Sbc-like galaxy SN~Ia rate \citep[see also][]{li2011b}.
Taking this number at face value as the overall SN~Ia rate, we find
the rate of deflagrations in hybrid CONe WDs to be on the order of 1
percent of the SN~Ia rate. This relative rate will increase, if one
considers galaxies with active star formation rather than older
stellar populations (like the Milky Way). Given the large
uncertainties in the observed rate of SNe~Iax (different authors give
values between 5 and 30 percent of the overall rate of SNe~Ia,
\citealt{li2011a, foley2013b, white2015a}), our estimated rate seems
in rough agreement with faint SNe~Iax.

\begin{figure}
  \centering
  \includegraphics[width=0.9\linewidth]{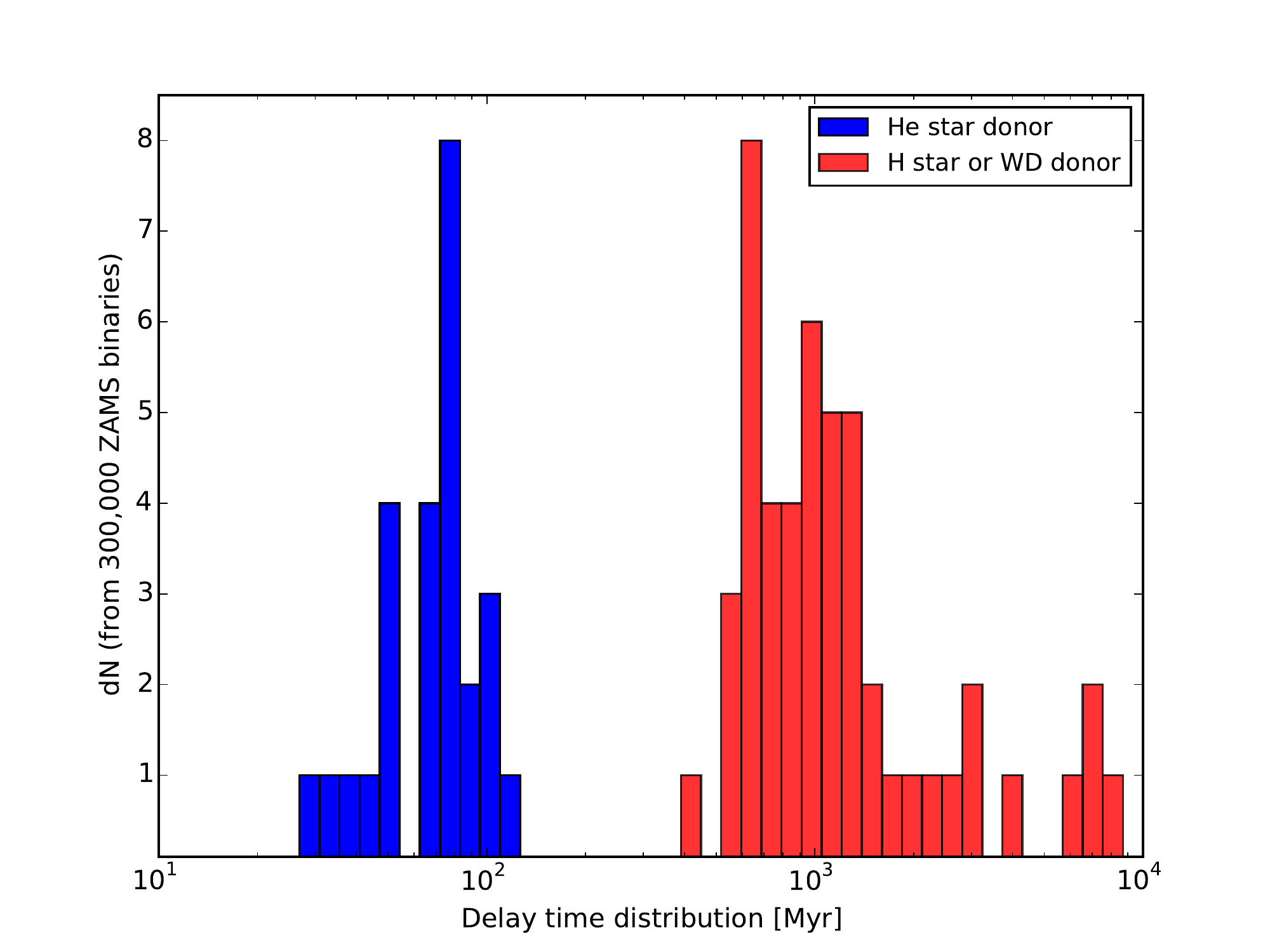}
  \caption{Delay time distribution of CONe WDs that approach \mch\ due
    to accretion from a binary star companion. Blue systems are those
    with helium-burning stars as donors while the red systems contain
    main-sequence, sub-giant, giant or WD donors. Numbers (y-axis) are
    not scaled to an absolute rate but rather represent the original
    numbers from our model.  An estimate of absolute rates (over a
    Hubble time) is given in the text.}
  \label{fig:dtd}
\end{figure}

In Figure~\ref{fig:dtd} we show the delay time distribution for the
population of hybrid CONe WDs estimated from our population synthesis
model. One third of our hybrid systems have delay times $< 150$\,Myr
with the shortest delay time occurring at 30\,Myr.  {\em All} of our
prompt systems below $150$\,Myr have helium-burning stars as donors
(shown in blue).  The range of delay times for these progenitors
agrees with the results of \citet{wang2014a}, who estimated the delay
time range for CONe hybrid WDs that accrete toward \mch\ from helium
stars to be 28 to 178\,Myr.  Our results are consistent with the fact
that SNe~Iax are found among young stellar populations; SN~2008ha is
estimated to have a delay time of $\lesssim 80$\,Myr
\citep{foley2014b}.  Nearly half of our hybrid systems, however, have
hydrogen-burning donors, while 17 percent have He WD donors (all shown
in red). All of these systems have longer delay times (shortest one is
420\,Myr).  The main factor that determines whether a progenitor will
have a very short delay time and a helium-burning star donor, or
whether it will have a slightly longer delay time with a
hydrogen-burning star (or WD) donor is the initial mass of the donor
star. For the prompt systems, the ZAMS mass of the donor star ranges
from $5-10$ \msun\ while for the rest, it ranges from $1-3$
\msun. Given our findings, it is unlikely that most progenitors of
faint SN~Iax events originate from systems with hydrogen-burning or WD
donors, though our results strongly support the idea that faint SN~Iax
events arise from CONe WDs that accrete from helium burning stars.

\section{Conclusion}
\label{sec:conclusion}

Previous models of deflagrations in near-\mch\ CO WDs had difficulties
in explaining faint SNe~Iax, since they produced significantly too
large amounts of \nuc{56}{Ni} \citep{fink2014a}. To reduce the
\nuc{56}{Ni} production, burning has to quench at an early
stage. Deflagrations in hybrid CONe WDs, which have recently been
predicted by stellar evolution models \citep{denissenkov2013c,
  chen2014a}, may lead to such a quenching when the burning front
transitions from the WD's CO-rich core to its ONe-rich mantle.

We have simulated an off-centre deflagration in a near-\mch\ hybrid
CONe WD. Assuming that the deflagration cannot propagate in the ONe
layer, we find that only 0.014\,\msun\ of material is ejected while
the remainder of the mass stays bound. The ejecta consist
predominantly of IGEs, O, C, Si and S. Containing only
$3.4\times10^{-3}$\,\msun\ of \nuc{56}{Ni}, the ejecta will give rise
to a faint transient.

Performing radiative transfer simulations for our model ejecta, we
find peak absolute magnitudes in the range $M_\mathrm{max}^B=-13.2$ to
$-14.6$\,mag depending on the viewing angle. This is in good agreement
with the observed peak brightness of the faint Type Iax SN~2008ha
($M_\mathrm{max}^B=-13.7$\,mag). Between peak and 10\,d thereafter we
also find reasonable agreement between the observed spectral shape of
SN~2008ha and our model. This shows for the first time that
deflagrations in near-\mch\ hybrid CONe WDs can lead to faint
transients with a display similar to faint SNe~Iax.

However, our model still has some shortcomings. At epochs before
maximum and later than 10\,d past maximum the spectral evolution is
too fast, probably indicating that the ejecta mass is too low. Since
the exact thermal and chemical structure of \mch\ hybrid CONe WDs is
not well constrained \citep{denissenkov2015a}, we used a simple model
for the progenitor. Future work exploring a range of different initial
conditions will show whether even better agreement with faint SNe~Iax
is possible from deflagrations in hybrid CONe WDs. More vigorous
ignition setups, similar to those explored in CO WDs by
\citet{fink2014a}, or hybrid WDs with more massive CO cores might also
be able to explain intermediate-luminosity SNe~Iax like SN~2007qd
\citep{mcclelland2010a}.

From binary population synthesis calculations we find that near-\mch\
hybrid CONe WDs that may produce faint SNe~Iax occur at a rate of 1
percent of the Galactic SN~Ia rate. The delay times of these systems
depend sensitively on the nature of the donor star. He-burning donors
(delay times 30 to 150\,Myr) are compatible with the estimated delay
time of SN~2008ha \citep[80\,Myr,][]{foley2014a} and the observed
preference for star-forming environment among SNe~Iax
\citep{lyman2013a}. H-burning and WD donors have somewhat longer delay
times.

Finally, hydrodynamic simulations with adaptive grid codes will be
needed to resolve the structure and thermal state of the bound remnant
predicted by our model. This will be crucial to assess the influence
of the bound remnant on the observational display of the explosion and
allow for a detailed comparison to the potential stellar remnant
detected at the site of the faint SN~2008ha. Following the long-term
evolution of the near-\mch\ bound remnant and companion star will show
whether subsequent mass-transfer episodes can lead to recurrent
explosions or accretion-induced collapse.

\section*{Acknowledgements}

This work was supported by the Deutsche Forschungs\-gemeinschaft via
the Transregional Collaborative Research Center TRR 33 `The Dark
Universe' and the Emmy Noether Program (RO 3676/1-1).  Parts of this
research were conducted by the Australian Research Council Centre of
Excellence for All-sky Astrophysics (CAASTRO) through project number
CE110001020 and by the the ARC Laureate Grant FL0992131.

AJR thanks A. Karakas and M. Childress for informative discussion. FKR
is supported by the ARCHES prize of the German Ministry of Education
and Research (BMBF).  KSM acknowledges financial support from the DFG
via the graduate school ``Theoretical Astrophysics and Particle
Physics'' at the University of W\"urzburg (GRK 1147). RP acknowledges
support by the European Research Council under ERC-StG EXAGAL-308037
and by the Klaus Tschira Foundation. SAS acknowledges support from
STFC grant ST/L000709/1. STO acknowledges financial support from
Studienstiftung des deutschen Volkes. We also thank the DAAD/Go8
German-Australian exchange programme for travel support.

The authors gratefully acknowledge the Gauss Centre for Supercomputing
(GCS) for providing computing time through the John von Neumann
Institute for Computing (NIC) on the GCS share of the supercomputer
JUQUEEN at J\"ulich Supercomputing Centre (JSC). GCS is the alliance
of the three national supercomputing centres HLRS (Universit\"at
Stuttgart), JSC (Forschungszentrum J\"ulich), and LRZ (Bayerische
Akademie der Wissenschaften), funded by the German Federal Ministry of
Education and Research (BMBF) and the German State Ministries for
Research of Baden-W\"urttemberg (MWK), Bayern (StMWFK) and
Nordrhein-Westfalen (MIWF).

\bibliographystyle{mn2e}

\begin{thebibliography}{57}
\expandafter\ifx\csname natexlab\endcsname\relax\def\natexlab#1{#1}\fi

\bibitem[{{Badenes} \& {Maoz}(2012)}]{badenes2012a}
{Badenes} C., {Maoz} D., 2012, \apjl, 749, L11

\bibitem[{{Belczynski} {et~al}\mbox{.}(2002){Belczynski}, {Kalogera}, \&
  {Bulik}}]{belczynski2002a}
{Belczynski} K., {Kalogera} V., {Bulik} T., 2002, \apj, 572, 407

\bibitem[{{Belczynski} {et~al}\mbox{.}(2008){Belczynski}, {Kalogera}, {Rasio},
  {Taam}, {Zezas}, {Bulik}, {Maccarone}, \& {Ivanova}}]{belczynski2008a}
{Belczynski} K., {Kalogera} V., {Rasio} F.~A., {Taam} R.~E., {Zezas} A.,
  {Bulik} T., {Maccarone} T.~J., {Ivanova} N., 2008, \apjs, 174, 223

\bibitem[{{Branch} {et~al}\mbox{.}(2004){Branch}, {Baron}, {Thomas}, {Kasen},
  {Li}, \& {Filippenko}}]{branch2004a}
{Branch} D., {Baron} E., {Thomas} R.~C., {Kasen} D., {Li} W., {Filippenko}
  A.~V., 2004, \pasp, 116, 903

\bibitem[{{Chen} {et~al}\mbox{.}(2014){Chen}, {Herwig}, {Denissenkov}, \&
  {Paxton}}]{chen2014a}
{Chen} M.~C., {Herwig} F., {Denissenkov} P.~A., {Paxton} B., 2014, \mnras, 440,
  1274

\bibitem[{{Chornock} {et~al}\mbox{.}(2006){Chornock}, {Filippenko}, {Branch},
  {Foley}, {Jha}, \& {Li}}]{chornock2006a}
{Chornock} R., {Filippenko} A.~V., {Branch} D., {Foley} R.~J., {Jha} S., {Li}
  W., 2006, \pasp, 118, 722

\bibitem[{{Denissenkov} {et~al}\mbox{.}(2013){Denissenkov}, {Herwig}, {Truran},
  \& {Paxton}}]{denissenkov2013c}
{Denissenkov} P.~A., {Herwig} F., {Truran} J.~W., {Paxton} B., 2013, \apj, 772,
  37

\bibitem[{{Denissenkov} {et~al}\mbox{.}(2015){Denissenkov}, {Truran}, {Herwig},
  {Jones}, {Paxton}, {Nomoto}, {Suzuki}, \& {Toki}}]{denissenkov2015a}
{Denissenkov} P.~A., {Truran} J.~W., {Herwig} F., {Jones} S., {Paxton} B.,
  {Nomoto} K., {Suzuki} T., {Toki} H., 2015, \mnras, 447, 2696

\bibitem[{{Fink} {et~al}\mbox{.}(2014){Fink}, {Kromer}, {Seitenzahl},
  {Ciaraldi-Schoolmann}, {R{\"o}pke}, {Sim}, {Pakmor}, {Ruiter}, \&
  {Hillebrandt}}]{fink2014a}
{Fink} M. {et~al.}, 2014, \mnras, 438, 1762

\bibitem[{{Foley} {et~al}\mbox{.}(2010){Foley}, {Brown}, {Rest}, {Challis},
  {Kirshner}, \& {Wood-Vasey}}]{foley2010a}
{Foley} R.~J., {Brown} P.~J., {Rest} A., {Challis} P.~J., {Kirshner} R.~P.,
  {Wood-Vasey} W.~M., 2010, \apjl, 708, L61

\bibitem[{{Foley} {et~al}\mbox{.}(2013){Foley}, {Challis}, {Chornock},
  {Ganeshalingam}, {Li}, {Marion}, {Morrell}, {Pignata}, {Stritzinger},
  {Silverman}, {Wang}, {Anderson}, {Filippenko}, {Freedman}, {Hamuy}, {Jha},
  {Kirshner}, {McCully}, {Persson}, {Phillips}, {Reichart}, \&
  {Soderberg}}]{foley2013b}
{Foley} R.~J. {et~al.}, 2013, \apj, 767, 57

\bibitem[{{Foley} {et~al}\mbox{.}(2009){Foley}, {Chornock}, {Filippenko},
  {Ganeshalingam}, {Kirshner}, {Li}, {Cenko}, {Challis}, {Friedman}, {Modjaz},
  {Silverman}, \& {Wood-Vasey}}]{foley2009a}
{Foley} R.~J. {et~al.}, 2009, \aj, 138, 376

\bibitem[{{Foley} {et~al}\mbox{.}(2014{\natexlab{a}}){Foley}, {Fox}, {McCully},
  {Phillips}, {Sand}, {Zheng}, {Challis}, {Filippenko}, {Folatelli},
  {Hillebrandt}, {Hsiao}, {Jha}, {Kirshner}, {Kromer}, {Marion}, {Nelson},
  {Pakmor}, {Pignata}, {R{\"o}pke}, {Seitenzahl}, {Silverman}, {Skrutskie}, \&
  {Stritzinger}}]{foley2014a}
{Foley} R.~J. {et~al.}, 2014{\natexlab{a}}, \mnras, 443, 2887

\bibitem[{{Foley} {et~al}\mbox{.}(2014{\natexlab{b}}){Foley}, {McCully}, {Jha},
  {Bildsten}, {Fong}, {Narayan}, {Rest}, \& {Stritzinger}}]{foley2014b}
{Foley} R.~J., {McCully} C., {Jha} S.~W., {Bildsten} L., {Fong} W.-f.,
  {Narayan} G., {Rest} A., {Stritzinger} M.~D., 2014{\natexlab{b}}, \apj, 792,
  29

\bibitem[{{Gall} {et~al}\mbox{.}(2012){Gall}, {Taubenberger}, {Kromer}, {Sim},
  {Benetti}, {Blanc}, {Elias-Rosa}, \& {Hillebrandt}}]{gall2012a}
{Gall} E.~E.~E., {Taubenberger} S., {Kromer} M., {Sim} S.~A., {Benetti} S.,
  {Blanc} G., {Elias-Rosa} N., {Hillebrandt} W., 2012, \mnras, 427, 994

\bibitem[{{Garcia-Berro} {et~al}\mbox{.}(1997){Garcia-Berro}, {Ritossa}, \&
  {Iben}}]{garcia-berro1997a}
{Garcia-Berro} E., {Ritossa} C., {Iben} I.~J., 1997, \apj, 485, 765

\bibitem[{{Hillebrandt} \& {Niemeyer}(2000)}]{hillebrandt2000a}
{Hillebrandt} W., {Niemeyer} J.~C., 2000, \araa, 38, 191

\bibitem[{{Jha} {et~al}\mbox{.}(2006){Jha}, {Kirshner}, {Challis}, {Garnavich},
  {Matheson}, {Soderberg}, {Graves}, {Hicken}, {Alves}, {Arce}, {Balog},
  {Barmby}, {Barton}, {Berlind}, {Bragg}, {Brice{\~n}o}, {Brown}, {Buckley},
  {Caldwell}, {Calkins}, {Carter}, {Concannon}, {Donnelly}, {Eriksen},
  {Fabricant}, {Falco}, {Fiore}, {Garcia}, {G{\'o}mez}, {Grogin}, {Groner},
  {Groot}, {Haisch}, {Hartmann}, {Hergenrother}, {Holman}, {Huchra},
  {Jayawardhana}, {Jerius}, {Kannappan}, {Kim}, {Kleyna}, {Kochanek},
  {Koranyi}, {Krockenberger}, {Lada}, {Luhman}, {Luu}, {Macri}, {Mader},
  {Mahdavi}, {Marengo}, {Marsden}, {McLeod}, {McNamara}, {Megeath}, {Moraru},
  {Mossman}, {Muench}, {Mu{\~n}oz}, {Muzerolle}, {Naranjo}, {Nelson-Patel},
  {Pahre}, {Patten}, {Peters}, {Peters}, {Raymond}, {Rines}, {Schild},
  {Sobczak}, {Spahr}, {Stauffer}, {Stefanik}, {Szentgyorgyi}, {Tollestrup},
  {V{\"a}is{\"a}nen}, {Vikhlinin}, {Wang}, {Willner}, {Wolk}, {Zajac}, {Zhao},
  \& {Stanek}}]{jha2006a}
{Jha} S. {et~al.}, 2006, \aj, 131, 527

\bibitem[{{Jordan} {et~al}\mbox{.}(2012){Jordan}, {Perets}, {Fisher}, \& {van
  Rossum}}]{jordan2012b}
{Jordan}, IV G.~C., {Perets} H.~B., {Fisher} R.~T., {van Rossum} D.~R., 2012,
  \apjl, 761, L23

\bibitem[{{Kromer} {et~al}\mbox{.}(2013){Kromer}, {Fink}, {Stanishev},
  {Taubenberger}, {Ciaraldi-Schoolman}, {Pakmor}, {R{\"o}pke}, {Ruiter},
  {Seitenzahl}, {Sim}, {Blanc}, {Elias-Rosa}, \& {Hillebrandt}}]{kromer2013a}
{Kromer} M. {et~al.}, 2013, \mnras, 429, 2287

\bibitem[{{Kromer} \& {Sim}(2009)}]{kromer2009a}
{Kromer} M., {Sim} S.~A., 2009, \mnras, 398, 1809

\bibitem[{{Kromer} {et~al}\mbox{.}(2010){Kromer}, {Sim}, {Fink}, {R{\"o}pke},
  {Seitenzahl}, \& {Hillebrandt}}]{kromer2010a}
{Kromer} M., {Sim} S.~A., {Fink} M., {R{\"o}pke} F.~K., {Seitenzahl} I.~R.,
  {Hillebrandt} W., 2010, \apj, 719, 1067

\bibitem[{{Li} {et~al}\mbox{.}(2011{\natexlab{a}}){Li}, {Bloom},
  {Podsiadlowski}, {Miller}, {Cenko}, {Jha}, {Sullivan}, {Howell}, {Nugent},
  {Butler}, {Ofek}, {Kasliwal}, {Richards}, {Stockton}, {Shih}, {Bildsten},
  {Shara}, {Bibby}, {Filippenko}, {Ganeshalingam}, {Silverman}, {Kulkarni},
  {Law}, {Poznanski}, {Quimby}, {McCully}, {Patel}, {Maguire}, \&
  {Shen}}]{li2011b}
{Li} W. {et~al.}, 2011{\natexlab{a}}, \nat, 480, 348

\bibitem[{{Li} {et~al}\mbox{.}(2003){Li}, {Filippenko}, {Chornock}, {Berger},
  {Berlind}, {Calkins}, {Challis}, {Fassnacht}, {Jha}, {Kirshner}, {Matheson},
  {Sargent}, {Simcoe}, {Smith}, \& {Squires}}]{li2003a}
{Li} W. {et~al.}, 2003, \pasp, 115, 453

\bibitem[{{Li} {et~al}\mbox{.}(2011{\natexlab{b}}){Li}, {Leaman}, {Chornock},
  {Filippenko}, {Poznanski}, {Ganeshalingam}, {Wang}, {Modjaz}, {Jha}, {Foley},
  \& {Smith}}]{li2011a}
{Li} W. {et~al.}, 2011{\natexlab{b}}, \mnras, 412, 1441

\bibitem[{{Liu} {et~al}\mbox{.}(2013){Liu}, {Kromer}, {Fink}, {Pakmor},
  {R{\"o}pke}, {Chen}, {Wang}, \& {Han}}]{liu2013c}
{Liu} Z.-W., {Kromer} M., {Fink} M., {Pakmor} R., {R{\"o}pke} F.~K., {Chen}
  X.~F., {Wang} B., {Han} Z.~W., 2013, \apj, 778, 121

\bibitem[{{Lyman} {et~al}\mbox{.}(2013){Lyman}, {James}, {Perets}, {Anderson},
  {Gal-Yam}, {Mazzali}, \& {Percival}}]{lyman2013a}
{Lyman} J.~D., {James} P.~A., {Perets} H.~B., {Anderson} J.~P., {Gal-Yam} A.,
  {Mazzali} P., {Percival} S.~M., 2013, \mnras, 434, 527

\bibitem[{{McClelland} {et~al}\mbox{.}(2010){McClelland}, {Garnavich},
  {Galbany}, {Miquel}, {Foley}, {Filippenko}, {Bassett}, {Wheeler}, {Goobar},
  {Jha}, {Sako}, {Frieman}, {Sollerman}, {Vinko}, \&
  {Schneider}}]{mcclelland2010a}
{McClelland} C.~M. {et~al.}, 2010, \apj, 720, 704

\bibitem[{{McCully} {et~al}\mbox{.}(2014){McCully}, {Jha}, {Foley}, {Bildsten},
  {Fong}, {Kirshner}, {Marion}, {Riess}, \& {Stritzinger}}]{mccully2014a}
{McCully} C. {et~al.}, 2014, \nat, 512, 54

\bibitem[{{Meng} \& {Podsiadlowski}(2014)}]{meng2014a}
{Meng} X., {Podsiadlowski} P., 2014, \apjl, 789, L45

\bibitem[{{Moriya} {et~al}\mbox{.}(2010){Moriya}, {Tominaga}, {Tanaka},
  {Nomoto}, {Sauer}, {Mazzali}, {Maeda}, \& {Suzuki}}]{moriya2010a}
{Moriya} T., {Tominaga} N., {Tanaka} M., {Nomoto} K., {Sauer} D.~N., {Mazzali}
  P.~A., {Maeda} K., {Suzuki} T., 2010, \apj, 719, 1445

\bibitem[{{Nomoto}(1984)}]{nomoto1984b}
{Nomoto} K., 1984, \apj, 277, 791

\bibitem[{{Nomoto}(1987)}]{nomoto1987a}
{Nomoto} K., 1987, \apj, 322, 206

\bibitem[{{Nomoto} \& {Kondo}(1991)}]{nomoto1991a}
{Nomoto} K., {Kondo} Y., 1991, \apjl, 367, L19

\bibitem[{{Osher} \& {Sethian}(1988)}]{osher1988a}
{Osher} S., {Sethian} J.~A., 1988, Journal of Computational Physics, 79, 12

\bibitem[{{Phillips} {et~al}\mbox{.}(2007){Phillips}, {Li}, {Frieman},
  {Blinnikov}, {DePoy}, {Prieto}, {Milne}, {Contreras}, {Folatelli}, {Morrell},
  {Hamuy}, {Suntzeff}, {Roth}, {Gonz{\'a}lez}, {Krzeminski}, {Filippenko},
  {Freedman}, {Chornock}, {Jha}, {Madore}, {Persson}, {Burns}, {Wyatt},
  {Murphy}, {Foley}, {Ganeshalingam}, {Serduke}, {Krisciunas}, {Bassett},
  {Becker}, {Dilday}, {Eastman}, {Garnavich}, {Holtzman}, {Kessler},
  {Lampeitl}, {Marriner}, {Frank}, {Marshall}, {Miknaitis}, {Sako},
  {Schneider}, {van der Heyden}, \& {Yasuda}}]{phillips2007a}
{Phillips} M.~M. {et~al.}, 2007, \pasp, 119, 360

\bibitem[{{Reinecke} {et~al}\mbox{.}(2002){Reinecke}, {Hillebrandt}, \&
  {Niemeyer}}]{reinecke2002b}
{Reinecke} M., {Hillebrandt} W., {Niemeyer} J.~C., 2002, \aap, 386, 936

\bibitem[{{Reinecke} {et~al}\mbox{.}(1999){Reinecke}, {Hillebrandt},
  {Niemeyer}, {Klein}, \& {Gr{\"o}bl}}]{reinecke1999a}
{Reinecke} M., {Hillebrandt} W., {Niemeyer} J.~C., {Klein} R., {Gr{\"o}bl} A.,
  1999, \aap, 347, 724

\bibitem[{{R{\"o}pke}(2005)}]{roepke2005c}
{R{\"o}pke} F.~K., 2005, \aap, 432, 969

\bibitem[{{R{\"o}pke} {et~al}\mbox{.}(2006){R{\"o}pke}, {Hillebrandt},
  {Niemeyer}, \& {Woosley}}]{roepke2006a}
{R{\"o}pke} F.~K., {Hillebrandt} W., {Niemeyer} J.~C., {Woosley} S.~E., 2006,
  \aap, 448, 1

\bibitem[{{R{\"o}pke} {et~al}\mbox{.}(2012){R{\"o}pke}, {Kromer}, {Seitenzahl},
  {Pakmor}, {Sim}, {Taubenberger}, {Ciaraldi-Schoolmann}, {Hillebrandt},
  {Aldering}, {Antilogus}, {Baltay}, {Benitez-Herrera}, {Bongard}, {Buton},
  {Canto}, {Cellier-Holzem}, {Childress}, {Chotard}, {Copin}, {Fakhouri},
  {Fink}, {Fouchez}, {Gangler}, {Guy}, {Hachinger}, {Hsiao}, {Chen},
  {Kerschhaggl}, {Kowalski}, {Nugent}, {Paech}, {Pain}, {Pecontal}, {Pereira},
  {Perlmutter}, {Rabinowitz}, {Rigault}, {Runge}, {Saunders}, {Smadja},
  {Suzuki}, {Tao}, {Thomas}, {Tilquin}, \& {Wu}}]{roepke2012a}
{R{\"o}pke} F.~K. {et~al.}, 2012, \apjl, 750, L19

\bibitem[{{Ruiter} {et~al}\mbox{.}(2014){Ruiter}, {Belczynski}, {Sim},
  {Seitenzahl}, \& {Kwiatkowski}}]{ruiter2014a}
{Ruiter} A.~J., {Belczynski} K., {Sim} S.~A., {Seitenzahl} I.~R., {Kwiatkowski}
  D., 2014, \mnras, 440, L101

\bibitem[{{Sahu} {et~al}\mbox{.}(2008){Sahu}, {Tanaka}, {Anupama}, {Kawabata},
  {Maeda}, {Tominaga}, {Nomoto}, {Mazzali}, \& {Prabhu}}]{sahu2008a}
{Sahu} D.~K. {et~al.}, 2008, \apj, 680, 580

\bibitem[{{Schmidt} {et~al}\mbox{.}(2006{\natexlab{a}}){Schmidt}, {Niemeyer},
  \& {Hillebrandt}}]{schmidt2006b}
{Schmidt} W., {Niemeyer} J.~C., {Hillebrandt} W., 2006{\natexlab{a}}, \aap,
  450, 265

\bibitem[{{Schmidt} {et~al}\mbox{.}(2006{\natexlab{b}}){Schmidt}, {Niemeyer},
  {Hillebrandt}, \& {R{\"o}pke}}]{schmidt2006c}
{Schmidt} W., {Niemeyer} J.~C., {Hillebrandt} W., {R{\"o}pke} F.~K.,
  2006{\natexlab{b}}, \aap, 450, 283

\bibitem[{{Seitenzahl} {et~al}\mbox{.}(2010){Seitenzahl}, {R{\"o}pke}, {Fink},
  \& {Pakmor}}]{seitenzahl2010a}
{Seitenzahl} I.~R., {R{\"o}pke} F.~K., {Fink} M., {Pakmor} R., 2010, \mnras,
  407, 2297

\bibitem[{{Siess}(2006)}]{siess2006a}
{Siess} L., 2006, \aap, 448, 717

\bibitem[{{Sim}(2007)}]{sim2007b}
{Sim} S.~A., 2007, \mnras, 375, 154

\bibitem[{{Sim} {et~al}\mbox{.}(2013){Sim}, {Seitenzahl}, {Kromer},
  {Ciaraldi-Schoolmann}, {R{\"o}pke}, {Fink}, {Hillebrandt}, {Pakmor},
  {Ruiter}, \& {Taubenberger}}]{sim2013a}
{Sim} S.~A. {et~al.}, 2013, \mnras, 436, 333

\bibitem[{{Smiljanovski} {et~al}\mbox{.}(1997){Smiljanovski}, {Moser}, \&
  {Klein}}]{smiljanovski1997a}
{Smiljanovski} V., {Moser} V., {Klein} R., 1997, Combustion Theory Modelling,
  1, 183

\bibitem[{{Stritzinger} {et~al}\mbox{.}(2014){Stritzinger}, {Hsiao}, {Valenti},
  {Taddia}, {Rivera-Thorsen}, {Leloudas}, {Maeda}, {Pastorello}, {Phillips},
  {Pignata}, {Baron}, {Burns}, {Contreras}, {Folatelli}, {Hamuy},
  {H{\"o}flich}, {Morrell}, {Prieto}, {Benetti}, {Campillay}, {Haislip},
  {LaClutze}, {Moore}, \& {Reichart}}]{stritzinger2014a}
{Stritzinger} M.~D. {et~al.}, 2014, \aap, 561, A146

\bibitem[{{Stritzinger} {et~al}\mbox{.}(2015){Stritzinger}, {Valenti},
  {Hoeflich}, {Baron}, {Phillips}, {Taddia}, {Foley}, {Hsiao}, {Jha},
  {McCully}, {Pandya}, {Simon}, {Benetti}, {Brown}, {Burns}, {Campillay},
  {Contreras}, {F{\"o}rster}, {Holmbo}, {Marion}, {Morrell}, \&
  {Pignata}}]{stritzinger2015a}
{Stritzinger} M.~D. {et~al.}, 2015, \aap, 573, A2

\bibitem[{{Timmes} \& {Woosley}(1992)}]{timmes1992a}
{Timmes} F.~X., {Woosley} S.~E., 1992, \apj, 396, 649

\bibitem[{{Travaglio} {et~al}\mbox{.}(2004){Travaglio}, {Hillebrandt},
  {Reinecke}, \& {Thielemann}}]{travaglio2004a}
{Travaglio} C., {Hillebrandt} W., {Reinecke} M., {Thielemann} F.-K., 2004,
  \aap, 425, 1029

\bibitem[{{Valenti} {et~al}\mbox{.}(2009){Valenti}, {Pastorello}, {Cappellaro},
  {Benetti}, {Mazzali}, {Manteca}, {Taubenberger}, {Elias-Rosa}, {Ferrando},
  {Harutyunyan}, {Hentunen}, {Nissinen}, {Pian}, {Turatto}, {Zampieri}, \&
  {Smartt}}]{valenti2009a}
{Valenti} S. {et~al.}, 2009, \nat, 459, 674

\bibitem[{{Wang} {et~al}\mbox{.}(2014){Wang}, {Meng}, {Liu}, {Liu}, \&
  {Han}}]{wang2014a}
{Wang} B., {Meng} X., {Liu} D.-D., {Liu} Z.-W., {Han} Z., 2014, \apjl, 794, L28

\bibitem[{{White} {et~al}\mbox{.}(2015){White}, {Kasliwal}, {Nugent},
  {Gal-Yam}, {Howell}, {Sullivan}, {Goobar}, {Piro}, {Bloom}, {Kulkarni},
  {Laher}, {Masci}, {Ofek}, {Surace}, {Ben-Ami}, {Cao}, {Cenko}, {Hook},
  {J{\"o}nsson}, {Matheson}, {Sternberg}, {Quimby}, \& {Yaron}}]{white2015a}
{White} C.~J. {et~al.}, 2015, \apj, 799, 52

\end{thebibliography}

\label{lastpage}
\end{document}